# Local Interstellar Neutral Hydrogen sampled in-situ by IBEX

Lukas Saul[1], Peter Wurz[1], Diego Rodriguez[1], Jürgen Scheer[1]

Eberhard Möbius[2], Nathan Schwadron[2], Harald Kucharek [2], Trevor Leonard[2],

Maciej Bzowski[3],

Stephen Fuselier[4],

Geoff Crew[5],

Dave McComas[6,7]

(1) University of Bern, Bern, Switzerland
(2) University of New Hampshire, Durham NH, USA
(3) Space Science Centre PAS, Warsaw, Poland
(4) Lockheed Martin Advanced Technology Center, Palo Alto CA, USA
(5) Haystack Observatory, Massachusetts Institute of Technology, MA, USA
(6) Southwest Research Institute, San Antonio TX, USA
(7) University of Texas at San Antonio, San Antonio, TX 78249, USA

**Abstract.** Hydrogen gas is the dominant component of the local interstellar medium. However, due to ionization and interaction with the heliosphere, direct sampling of neutral hydrogen in the inner heliosphere is more difficult than sampling the local interstellar neutral helium, which penetrates deep into the heliosphere. In this paper we report on the first detailed analysis of the direct sampling of neutral hydrogen from the local interstellar medium. We confirm that the arrival direction of hydrogen is offset from that of the local Helium component. We further report the discovery of a variation of the penetrating Hydrogen over the first two years of IBEX observations. Observations are consistent with hydrogen experiencing an effective ratio of outward solar radiation pressure to inward gravitational force greater than unity ($\mu$>1); the temporal change observed in the local interstellar hydrogen flux can be explained with solar variability.

## 1. Interstellar Wind

Most of our knowledge of the local interstellar medium comes from line-of-sight spectroscopy of nearby stars. The Sun resides in a local bubble of hot diffuse gas, in a larger system of interstellar clouds [Redfield et al., 2009; Frisch et al., 2002]. This gas interacts with the solar environment affecting the shape and structure of the heliosphere and cosmic ray intensities at Earth [e.g., Müller et al., 2006]. This interaction is controlled largely by bulk properties of the very local interstellar material. It is these bulk properties, in particular those of the most common element, hydrogen, that we focus our attention on in this paper.

In addition to spectroscopic absorption studies, more direct measurements of this material have been made, though of course only in the location of our stellar system. Solar radiation backscatters from the neutral component of the interstellar gas, which has been observed for hydrogen [Lallement et al., 2005] and helium [Vallerga et al., 2004]. It is the very local interstellar medium and magnetic field that combined with the solar wind determine the size and shape of the heliosphere [Baranov et al., 1981; Zank & Müller, 2003; Izmodenov et al., 2003]. For this reason as well as for our understanding of the solar environment and interstellar matter, it is important to carefully observe interstellar material that enters the heliosphere and determine its physical parameters.

Due to ionization processes, most of the neutral interstellar wind does not survive into the inner heliosphere. As a result, the composition of residual wind at 1AU (where IBEX resides) is mostly Helium, since He is hardly affected at the heliospheric interface and ionization in the inner heliosphere is much less than for hydrogen. Measurements of helium neutral atoms by IBEX give us our best determination of physical parameters of the local interstellar medium (See [Möbius et al., 2011; Lee et al., 2011; Bzowski et al., 2011] in this issue). Further, it is usually assumed that due to the vast distances between stars the kinetic properties and charge states of all species in the LISM will reach an equilibrium, suggesting that the primary component of the neutral hydrogen will have the same bulk parameters ($T_\infty$, $V_\infty$) as helium. If this is correct, then a measurement of the difference between neutral H and He components of neutral LISM is an effective probe of the filtration of local interstellar gas by the heliosphere. The first direct observations of neutral hydrogen were made by IBEX shortly after launch and reported by Möbius et al. [2009]; in this study we extend those observations and make the first detailed quantitative analysis of IBEX's hydrogen results.

## 1.1 Ionization Processes and Filtration

When interstellar neutral atoms enter the sphere of influence of the Sun, they are subjected to ionization processes. This ionization produces an additional source population within the solar wind via generation of pickup ions. These pickup ions have been sampled by spacecraft instrumentation [Moebius et al., 1985] and change properties of the solar wind plasma [e.g. Isenberg et al., 2003] with important consequences in the outer heliosphere. However, for consideration of pristine interstellar neutrals, ionization is simply a loss process. Interstellar neutrals are ionized by UV light, charge exchange, and electron impact ionization [for review see Rucinski, 1996, Möbius et al., 2004]. Elements with lower ionization energies become depleted further out in the heliosphere. For the case of hydrogen most of the neutrals have been ionized as they reach 1AU and so prior to IBEX [Moebius et al., 2009] pickup ions of H from the LISM had been observed in situ only at 5 AU [Gloeckler et al., 1993] and from 6.4 to 8.2 AU as Cassini's trajectory carried it through the downstream direction and made the first in situ observations of an "interstellar hydrogen shadow" from these losses [McComas et al., 2004].

Helium as a noble gas has a high ionization potential and the dominant ionization process is UV excitation. The high relative ionization rates of H and He also mean that the neutral interstellar wind at the position of the Earth is predominantly helium. For the case of hydrogen, the precise UV ionization rate depends on the time-variable solar UV emission. Further, the charge exchange ionization rate with solar wind protons, which is much larger than the photo-ionization rate, depends on the solar wind conditions. This complication indicates that a good model of the hydrogen ionization rate must be time dependent, anisotropic (latitude dependent) as well as energy dependent. It is outside the scope of this paper to develop and use such a detailed model. For a simple model, we use an ionization rate that is inversely proportional to the square of the heliocentric distance. This assumption is much more close to reality for the case of helium, which is dominated by UV ionization.

Solar UV is also responsible for a radiation pressure which has important effects especially on lighter elements. In particular our observations are largely explained by the action of solar radiation pressure on Hydrogen, which is larger than the gravitational force [McComas et al., 2004; Bzowski et al., 2008].

## 2. The IBEX-LO Sensor

The IBEX satellite has two ENA sensors covering two overlapping energy ranges, designed for heliospheric remote sensing [McComas et al., 2009a]. The interstellar neutral wind is in the energy range of the IBEX-Lo sensor [Fuselier et al., 2009a; Möbius et al., 2009]. Interstellar neutrals are the brightest signal in the sky (apart from the Earth) observed in this energy range. The backgrounds present in the IBEX-Lo measurements from various sources are well below the signal of the interstellar neutrals [Wurz et al., 2009].

For a more thorough description of IBEX-Lo see [Fuselier et al., 2009b]. Here we describe briefly the sensor's detection capabilities in context of measuring the neutral interstellar wind. An entrance collimator rejects charged particles and limits the incoming particles in angular space, partially determining the angular resolution. Incoming neutrals then scatter off the Carbon Vapor Deposited diamond charge conversion surface, either gaining a negative charge or sputtering a negative ion from the surface [Scheer et al., 2006]. If neither charge conversion nor sputtering occurs upon incidence on the conversion surface, the particle will not be detected. For helium, the negative ion state is only metastable and conversion takes place with a probability $<10^{-5}$ [Wurz et al., 2008]. For this paper, we use the sputtered components to identify the helium atoms in the interstellar wind.

After the conversion surface, a negative ion with the appropriate energy will continue through an electrostatic energy analyzer and be post-accelerated into a Time-Of-Flight (TOF) mass spectrometer for coincidence detection and mass determination. For interstellar hydrogen it is charge conversion that enables our detection, and the TOF or mass per charge spectrum enables us to separate it from the sputtered species.

In addition to having been designed to maximize the detection efficiency, the IBEX-Lo sensor has a number of innovations that optimize signal to noise ratio and reduce background count rate. The TOF unit uses a triple coincidence system that measures time of flight three ways to verify the detection [Moebius et al., 2008]. For this paper we only use fully qualified triple coincidence counts. The times of flight are required to pass a checksum test to assure consistency with a particle moving through the sensor.

### 2.1) Viewing Geometry

As the IBEX spacecraft travels in orbit about the Earth as Earth orbits the Sun, its orientation is adjusted every orbit to maintain its attitude as an approximately sun-pointed spinner. This forces the sensors to have a field-of-view roughly perpendicular to the spacecraft-sun line and also to keep the solar panels properly oriented. With this geometry, there are thus two periods of the year in which interstellar neutral wind has a high probability of entering the IBEX-Lo sensor [Möbius et al., 2008; McComas et al., 2009a]. During one of these observation times, the motion of the Earth and thus the motion of the IBEX spacecraft is anti-parallel to that of the interstellar neutrals, which means the recorded interstellar atoms have an increased energy and fluence in the IBEX reference frame. We refer to this observation time as the "spring passage", and the other time period in which the Earth moves away from the neutral wind as the "fall passage". Due to the greatly reduced efficiency of detection and lower flux into the instrument during the fall passage, this time period has not yet provided a clear signal of the interstellar wind. A fall measurement of the interstellar beam would greatly increase the accuracy of interstellar parameters by eliminating degeneracies between these parameters implicit in producing the spring signal alone [see e.g. Bzowski et al., Moebius et al., Lee et al., this issue]. For Hydrogen however, we must rely on only the spring passage for our analysis.

The exact position in Earths orbit (day of year) in which IBEX observes the peak interstellar flux depends not only on the direction of the interstellar wind but also on the ratio of radiation pressure to the gravitational force: μ. A cartoon showing the viewing geometry for the spring passage using different values of μ is provided in Figure 1.

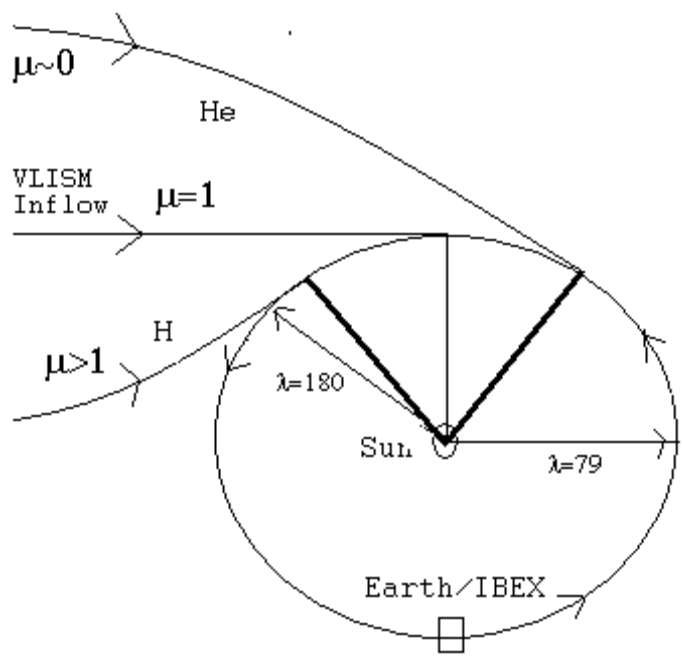

**Figure 1.** A cartoon showing how interstellar wind originating from the same direction will be observed at different times by IBEX depending on the ratio μ of radiation pressure to gravitational force (not to scale). The Earth at Vernal Equinox and inflow vector at ecliptic longitudes 180 and 79 respectively are shown for reference.

Observed longitudes of peak flows of H and He (dark lines) are given in Table 1.

Due to spin axis correction maneuvers the observed time series of particle flux will have a peak structure as IBEX enters the optimal location for observing interstellar neutrals. The structure of this peak is shown for the lowest energy channel (<E>=14eV) of IBEX-Lo in Fig. 2. The orientation of the spin axis of the spacecraft points the sensor into or away from the interstellar wind, creating the sawtooth structure. For comparison we show the results of a model which includes the exact position and orientation of the spacecraft in the helium wind (using physical parameters in [Möbius et al., 2004]), which shows good agreement. For a more precise comparison to interstellar helium model see [Möbius et al., 2011; Bzowski et al., 2011] this issue. The peak in Fig. 2 is almost entirely due to interstellar neutral helium, thus for this work we must subtract this component from the dataset to isolate the interstellar neutral hydrogen.

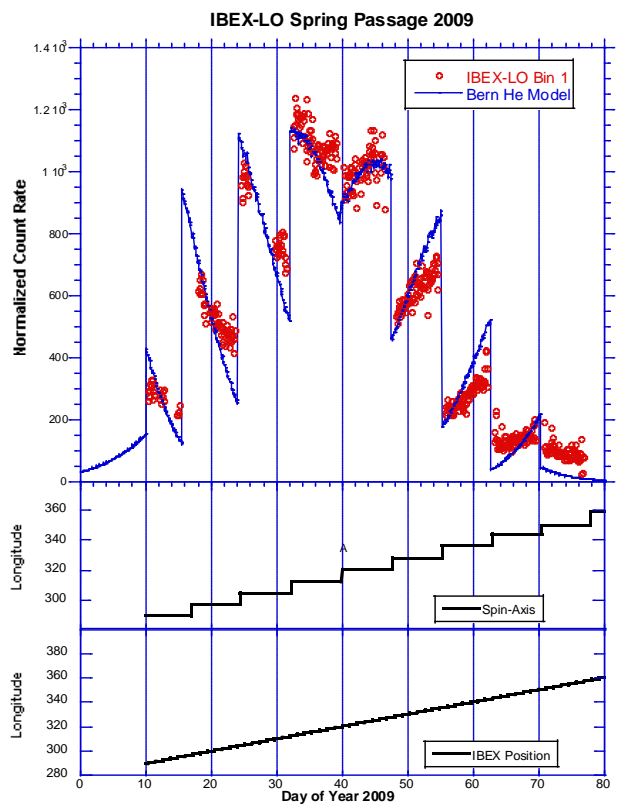

**Figure 2.** The count rate from IBEX-Lo energy bin 1 (center energy 14eV) integrated over the full azimuthal range is shown in red for the first spring passage (2009). The lower panel shows heliospheric position (longitude J2000) of the sun in the IBEX field of view and the middle panel shows pointing longitude of the spin-axis. For comparison a model of the IBEX helium observations using the Bern reverse neutral trajectory code and Witte parameters is shown in blue.

## 3. Subtraction of the Helium Component

The low energy neutral to ion surface conversion mechanism used by IBEX-Lo has one important drawback when it comes to composition analysis. Namely, sputtering of a hydrogen atom from the conversion surface by a more energetic particle produces a background in the observations. However, due to the different statistical properties of these separate particle distributions, for a large number of detections, we can subtract off the sputtered component using statistical methods. Here, this subtraction is done in two ways:

1. Determination and subtraction of typical sputtered TOF mass spectrum of interstellar helium atoms for selected time periods or spin phases. Hereafter, we refer to this method as the "subtraction method". During a given period of time we form a single observed spectrum. We then model this spectrum as the sum of a component produced by interstellar He and a smaller residual component produced by H, which we are solving for (see Fig. 4)

2. Energy analysis to separate H from the He component that peaks at higher energies due to its higher mass. In this case, we utilize the fact that the He component is predominantly observed at a higher energy step (ESA step 3) in the IBEX-Lo sensor. This energy step is then used to calibrate the differential energy flux of the He component, and can be more readily subtracted from the residual H component in ESA bin 1.

.

Both of these techniques suffer from the fact that the background we are trying to subtract (the He signal) is substantially larger than the signal we are looking for. This situation can compound small errors in the analysis. This drawback is partially mitigated by the fact that the signals do not occur at the same time, the same energy range, or the same TOF profile. *The combination of these different techniques, as well as different reference TOF spectra for the sputtered component are therefore used to make the identification robust.*

Critical in the analysis is an extensive set of calibration data obtained with the flight instrument in the MEFISTO laboratory at the University of Bern [Wieser & Wurz, 2005; Marti et al., 2000]. This includes the use of both a hydrogen and helium neutral beam at energies close to that expected for the interstellar component, to which we can compare the flight data.

### 3.1) TOF Spectrum Subtraction of Sputtered Component

A reference spectrum showing the sputtered component from neutral interstellar helium is shown in Fig. 3. As a test, we compared the results of three reference spectra when performing this analysis, one from calibration data (helium beam at expected interstellar energy); one from flight data during the peak flux of the interstellar helium (Fig. 3); and one from two orbits before the peak of the helium flux where there should be even less neutral interstellar hydrogen. Because of the different radiation pressure, ($\mu_{He} \sim 0$, $\mu_H \sim 1.2$ in 2009 and 2010) significant differences in the location of the respective maxima flux are expected.

We use only a single time of flight channel (TOF2) for subtracting the sputtered component, and analyze only triple coincidence particle events. The three reference spectra when used with the subtraction method returned similar values for the position of the hydrogen peak. However they displayed slightly different amounts of hydrogen away from the H peak (not shown). Because we focus in this paper on the position of the H peak, we use as a sputtered reference spectrum the in-flight data obtained during the peak He passage of orbit 64 (Fig. 4).

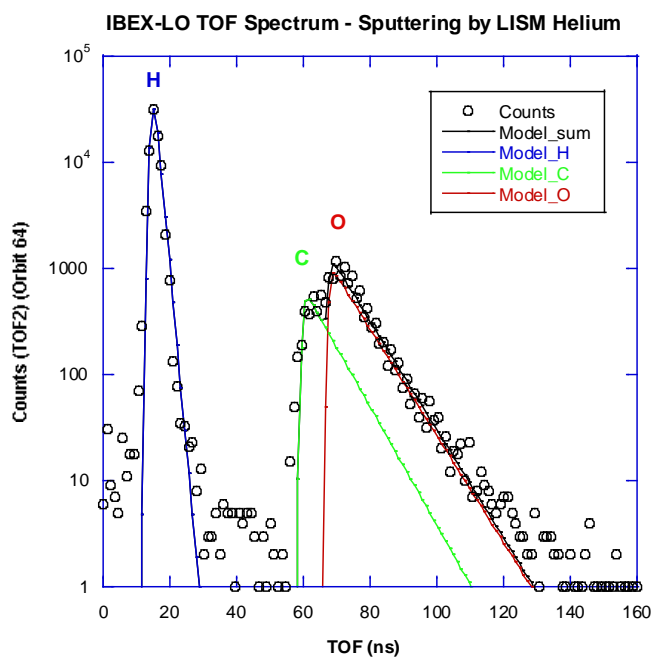

**Figure 3.** Sputtered spectrum in the lowest energy bin is shown for the TOF2 channel at maximum He flux (orbit 64). The sputtered peaks of hydrogen, Carbon, and Oxygen from the conversion surface show spreading due to energy straggling and angular scattering in the carbon foils in the TOF detector.

To subtract the sputtered component from a given measured TOF spectrum, we use a three part algorithm. We consider first only the secondary peaks (excluding H) of C and O in the measured spectrum and, using a single multiplicative scaling parameter we minimize the difference between the measured and reference spectra, apply a least squared method. Then, this optimal scaling parameter is applied to the sputtered H peak in the reference spectra, to determine the level of H in the spectra expected for the sputtered component. Finally the resulting full TOF spectra can be subtracted. The results of this procedure for two contrasting measured spectra are shown in Fig. 4, one taken from an orbit in which the signal is mostly He, and one taken from an orbit in which the spectrum is mostly H.

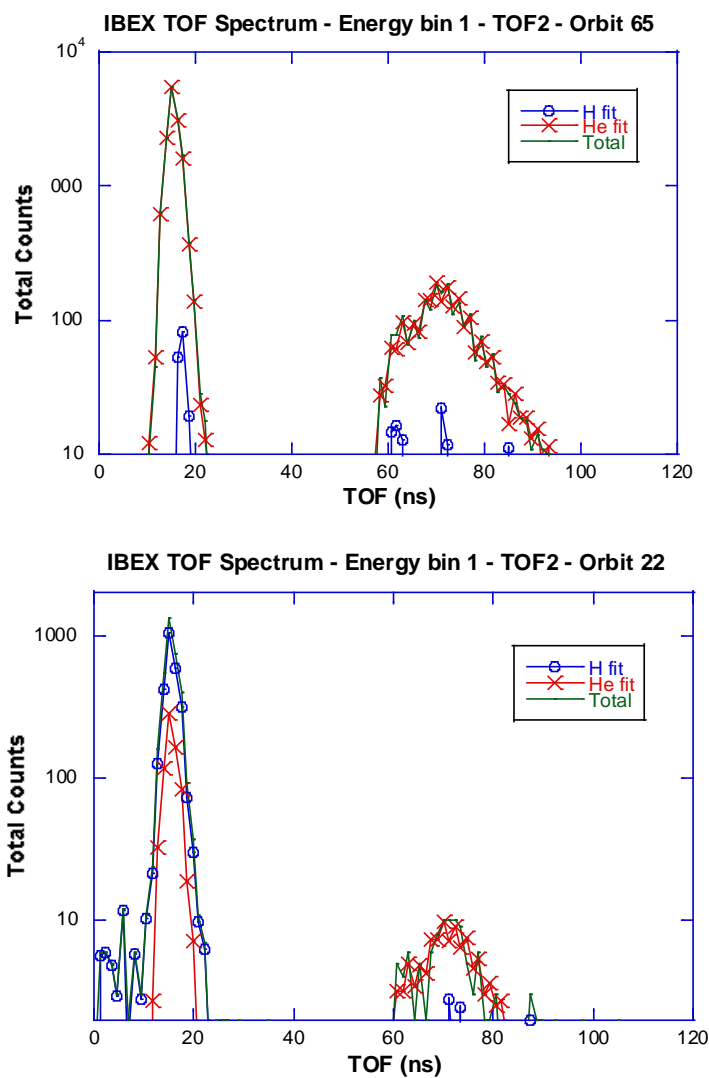

**Figure 4.** Two examples are shown illustrating the results of the algorithm used to subtract the sputtered component of helium. For the upper panel, the original spectrum (total) is found to be mostly helium, while for the lower plot the hydrogen peak is dominated by real converted neutral hydrogen. Note that the C and O peaks are always dominated by the He (sputtered) component.

We used this algorithm on the spectra obtained from the interstellar beam in each orbit, including only time periods which passed rigorous quality checks. These "select ISM flow observation times" agreed on by the IBEX team don't include anomalous dropouts or additions due to background sources and avoid saturation effects that are present during some time periods. The resulting per orbit time series is shown in Figure 5 for the spring passages of 2009 and 2010. The maximum real converted H count rate is less than 10% of the He (sputtered) rate and occurs over a month later. While the helium beam is stable in position, width, and amplitude from year to year, we observe a variation in the H component.

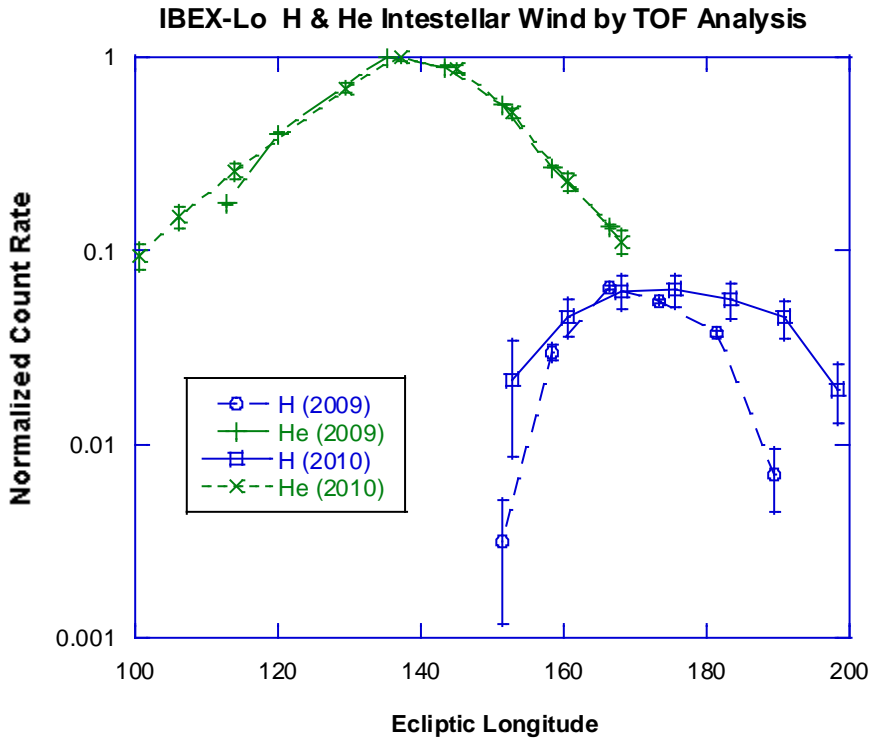

**IBEX-Lo  H & He Intestellar Wind by TOF Analysis**

**Figure 5.** The He and H components are separated on a per-orbit basis using the described algorithm on the TOF spectra. The first two spring passes of IBEX are shown. Error bars show statistical errors due to number of counts determined to be from the species in the corresponding orbit. Position is given in ecliptic longitude.

We can also use this method to isolate the H component of the IBEX count rate when separating the data not by orbit but by other parameters such as spin phase, i.e. position on the sky (elevation angle). This enables us to compare the heliospheric latitude of the He signal with that of the H signal. The IBEX viewing geometry means that we can scan the latitude every spin. However, the H count rates are too low to use this technique on shorter time periods or resolution higher than 2 degrees. This analysis shows a consistent offset in latitude between the He and H peaks during some orbits, as shown in the examples in Fig. 6 from the spring passage of 2010.

Because the LISM inflow direction is close to the ecliptic plane, the latitude of the observed peak is simpler to interpret as it is less dependent on ionization rate, radiation pressure to gravity ratio, or other potentially model dependent effects. However, the relative motion of the spacecraft produces a strong aberration which must be considered. These effects combine to produce an observed center latitude of the helium peak which varies with time (see [Bzowsi et al.; Möbius et al., Lee et al., this issue] for details). Further, during some orbits the observed shift in latitude between H and He components appears lower than others, with two extreme cases chosen for figure 6 with a 5 degree offset and 2 degree offset, respectively. A five degree offset is in agreement with the measurement of an H to He offset from backscattered solar radiation [Lallement et al., 2005]. The IBEX-Lo sensor has a +/- 6.5 degree field of view so the elevation angle (latitude) is oversampled. We report here the average variation between H and He as seen from IBEX, and for this work do not draw model-dependent conclusions on the relative offset between these beams further out in the heliosphere. At 1 AU our average observed offset is -2 degrees +/- 6.5 degrees, which is also consistent with no offset. Figure 6 also shows the substantial increase in H flux above the background. For analysis of the hydrogen signal far from the peak, the contribution of this background of heliospheric ENAs will be significant.

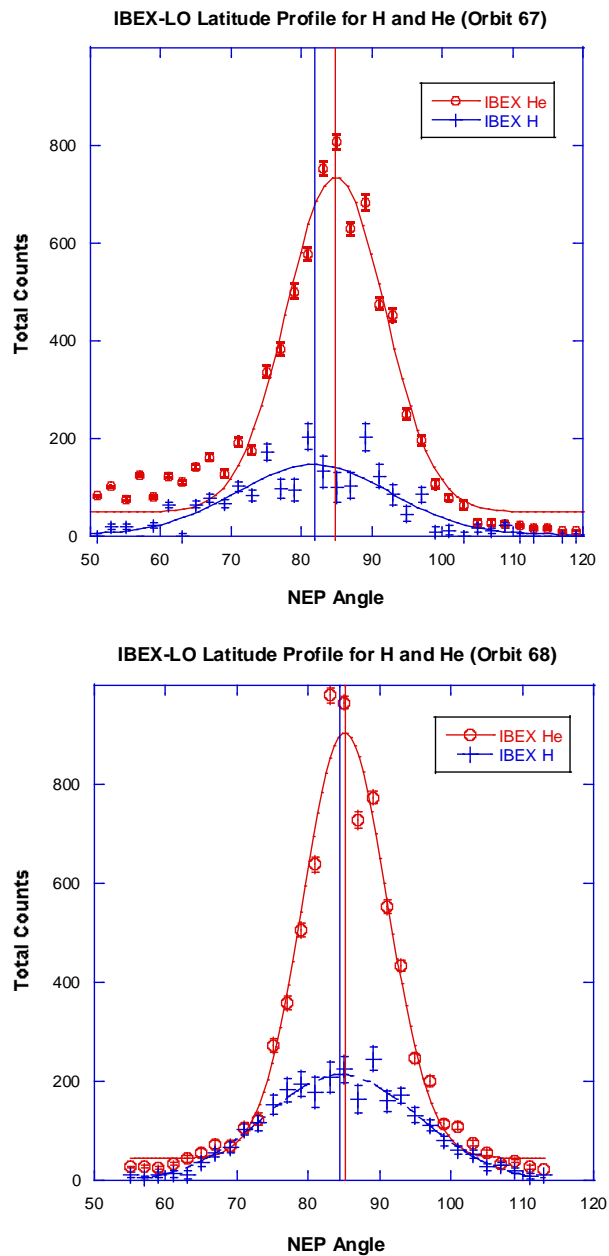

**Figure 6.** The helium and hydrogen components are separated on a directional (spin phase) basis using the described algorithm for the TOF spectra (upper plot) and energy analysis (lower plot). The angle is given in degrees from the north ecliptic pole, and the northward offset of the source from H with respect to He is clearly visible. The solid lines are Gaussian fits to the respective peaks, and their centroids are shown as vertical lines. Error bars show statistical errors.

### 3.2) Energy Analysis

The local interstellar He during the spring passage is expected to have a mean energy at 1AU of ~150eV, while for LISM H we expect a mean energy of ~15eV. It is the higher energy helium that produces the sputtered component also detected in the lowest energy bin. However, this difference in energy also gives us a further proxy to measure the pure LISM H. We expect the He to produce some sputtered signal in energy bin 3 (center energy 54eV) while there will be no directly converted LISM H in this channel. Thus, we can use the ratio of counts in bin 1 to bin 3 as a proxy for the H component. More precisely, we can make the assumption that the count rates in energy channels 1 and 3 are due to the H and He signals with certain efficiencies. i.e.:

$$R_1 = \alpha J_{He} + \beta J_H$$

$$R_3 = \gamma J_{He}$$

Where $\alpha, \beta, \gamma$ are the respective efficiencies (which incorporate geometric factors), $R_1$ and $R_3$ are the rates in energy channels 1 and 3, and $J_{He}$, $J_H$ are the differential energy fluxes of helium and hydrogen incident on the IBEX-Lo sensor. We can then solve for the real flux of hydrogen as:

$$J_H = \frac{1}{\beta}\left(R_1 - \frac{\alpha}{\gamma}R_3\right)$$

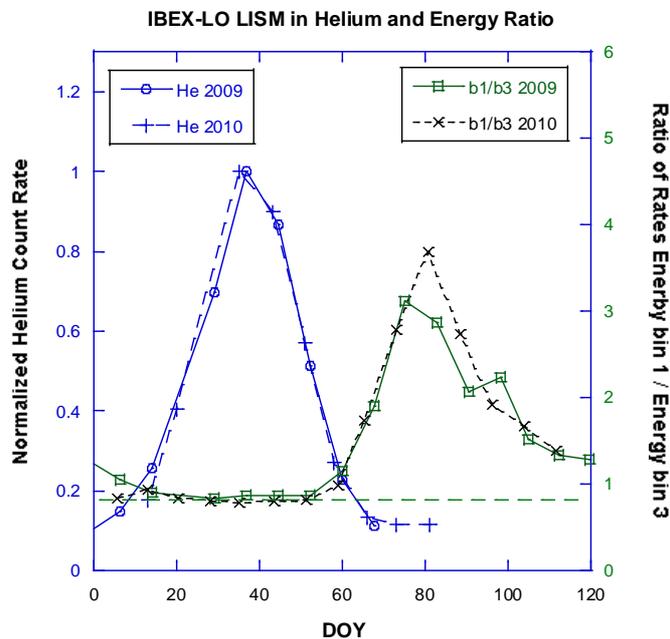

**Figure 7.** The ratio of counts in energy bins 1 and 3 is shown in comparison to the position of the interstellar helium wind peak. The energy ratio is shown on the right hand scale while the helium count rate is on the left hand scale. A dotted line indicates the value 0.8 used in the energy analysis characteristic of the interstellar helium peak.

During times when the signal is dominated by the He flux we can solve for $\frac{\alpha}{\gamma} = \frac{R_1}{R_3}$ . During the peak interstellar helium wind this number was quite stable at ~0.8, as shown in Fig. 7, which agrees well with IBEX-Lo laboratory calibration data. Thus, we have a proxy for the relative amount of H flux without needing to know precisely the efficiency β. The results of this analysis when applied to the per-orbit data during appropriate time periods are shown in Fig. 8. This simple analysis makes several assumptions which could produce some systematic error in the results. For example it does not allow for other species than H or He in the signal, or for a variation in ratio of energies for a single species even though the peak energy of a single species does vary somewhat with observation time due to the geometry. Sputtering efficiency is a steep function of energy in this regime. Further, we include a ~10% error in our estimate of the ratio in the two energy bins of $\frac{\alpha}{\gamma} = 0.8$ +/- 0.08.

The position of the H wind peak found using the described energy analysis disagrees somewhat with the results from the mass spectrum analysis, by from 1º to 2º in longitude. However the observation time of 8 days per data point suggests that they agree within uncertainties. The two methods do agree on two key observations:

1) The position of the H peak on the sky as measured at 1AU is substantially offset from the He peak (by ~35 degrees in longitude, see Table 1).

2) IBEX observed the peak H flow slightly later in the year during 2010 when compared to 2009, suggesting a higher value of $\mu$ during 2010 due to rising solar activity (see Table 1).

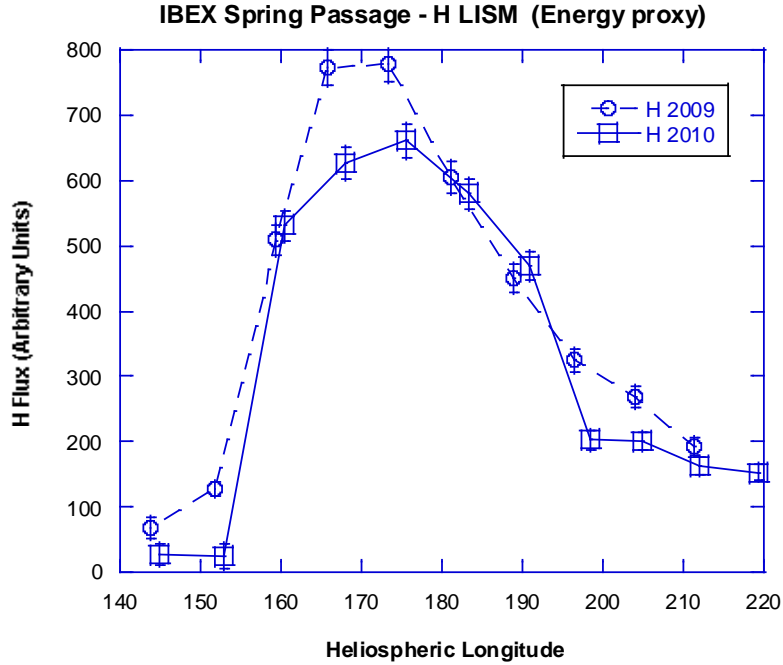

**Figure 8.** H components are shown on a per-orbit basis using the energy analysis. The slight shift of the peak to larger heliospheric longitude and broadening are consistent with the result from the TOF mass spectrum analysis. Error bars show statistical errors from determined number of H counts in each orbit.

## 4. Comparison to Helium Peak Location

The IBEX-Lo measurements unambiguously measure interstellar neutral H, and further show a clear offset from the He flow vector. In this study, we do not attempt to use these results to infer their source population in the outer heliosphere or beyond with a full heliospheric neutral trajectory model, but simply report on the directions and difference between the two components as observed at 1AU, here shown in Table 1. For helium, the use of such a model to determine local interstellar parameters outside the heliosphere is reported in this issue [Lee et al., 2011; Moebius et al., 2011; Bzowski et al., 2011]. However, difficulties present in modeling the H component are more complex than for helium. Therefore we present here measurement at 1AU, in terms of differences between the two components, as this is a model-independent result.

| | Peak $\lambda$ | FOV Center | Peak Norm. |
|---|---|---|---|
| He | 137°+-3° | 42.8° | 47°+-3° |
| H (tof)(2009) | 170°+-3° | 73.8° | 80°+-3° |
| H (tof)(2010) | 175°+-3° | 82.2° | 85°+-3° |
| H (e) (2009) | 172°+-3° | 80.1° | 82°+-3° |
| H (e) (2010) | 176°+-3° | 82.2° | 86°+-3° |

**Table 1**. Comparison of Longitudinal (J2000) He and H Interstellar Wind flow vectors as found from Gaussian fits to the data. Results from the analysis using the mass spectrum (tof) and results from energy analysis (e) are listed separately. Column 1 gives the ecliptic longitude of IBEX at the peak signal; Column 2 is the fixed field of view of the instrument during the 8 day orbit

containing the peak signal, while Column 3 is our result of H flow vector longitude normal to Earth-Sun line at peak flux.

Examining these results for the longitude of the peak interstellar flux, the most obvious difference is that the flow vector longitude is significantly larger than those reported by [Moebius et al., 2011] or by [Witte et al., 2004]. The parameter to explain this discrepancy is the ratio of radiation pressure to gravitation $\mu$. For an exact balance of gravitation and radiation pressure these flow vectors should match. Our observations of the Hydrogen peak are close to the predictions of [Bzowski et al., this issue] using the latest numbers of Lyman-alpha solar flux to determine $\mu$.

Also apparent from the measurements is the shift of the H peak from 2009 to 2010, consistent between both methods of H flux determination described here with the given temporal resolution. Our temporal resolution is limited by our use of data averaged over an entire orbit as well as any systematic errors in the H determination analysis. Although the pointing data and statistics are adequate enough to obtain far higher resolution [Möbius et al., Bzowski et al., this issue], we report a conservative error bar here due to the technique of extracting the hydrogen signal and its lower magnitude than that of the helium signal. As a crude maximum error, a ~7 day accumulation time as an effective FWHM of our measurement would suggest a standard deviation of ~3 days as a 1 sigma error to the day of year of the observed peak in Table 1. This is consistent with the difference observed between the two methods of hydrogen determination. We stress that at this point no model of the incoming interstellar hydrogen has been invoked and so a full error analysis of parameter space has not been performed.

## 5. Discussion of Results

IBEX-Lo measurements unambiguously show the detection of interstellar neutral H, and further show a clear offset from the He component. The offset in latitude southward is consistent with that observed by [Lallement et al., 2005], however our resolution at this stage of the analysis is not high enough to rigorously rule out a zero offset in latitude. However the offset in longitude of the 1 AU observations is much larger. This offset in longitude is likely due to a larger $\mu$ factor, representing the ratio of radiation pressure to gravitational force. This explanation of $\mu > 1$ is consistent with the small temporal variation of the offset, as the solar radiation has increased with the rising solar cycle over the last year.

In addition to further modeling of the neutral H component into the heliosphere, more work is needed to improve the statistics and refine our measurements of the exact position of the H peak. While our efforts in this paper focused on analysis of single IBEX orbits at one time, additional progress will be made by considering higher time resolution data. Further analysis of other TOF channels will also improve our fitting and He subtraction techniques, and this work is ongoing. These results should also be considered by global heliospheric models as an additional constrain based on the filtration of neutral hydrogen.

We do not report here on the density of neutral H as determined from IBEX measurements, as this quantity is strongly dependent on the ionization rate which decimates the incoming neutral gas. However, current modeling efforts by the IBEX team [see Bzowski et al., this issue] are proceeding to make this possible in the future.

**6. Acknowledgments**

The entire IBEX team is acknowledged for their great work and dedication to this successful mission. The financial support by the Swiss National Foundation and from the Polish Ministry for Science and Higher Education grants NS-1260-11-09 and N-N203-513-038 is also acknowledged. Helpful suggestions of Peter Bochsler and Fritz Bühler are acknowledged, as well as the work of the University of Bern laboratory staff including Georg Bodmer, Martin-Diego Busch and others. This work was supported by the IBEX mission as a part of NASA's Explorer Program.